\documentclass[pdflatex,sn-basic,iicol]{sn-jnl}

\usepackage{graphicx}%
\usepackage{multirow}%
\usepackage{amsmath,amssymb,amsfonts}%
\usepackage{amsthm}%
\usepackage{mathrsfs}%
\usepackage[title]{appendix}%
\usepackage{xcolor}%
\usepackage{textcomp}%
\usepackage{manyfoot}%
\usepackage{booktabs}%
\usepackage{algorithm}%
\usepackage{algorithmicx}%
\usepackage{algpseudocode}%
\usepackage{listings}%
\usepackage[T1]{fontenc}%

\begin{document}

\title[Nanohertz GW Astronomy]{The Dawn of Gravitational Wave Astronomy at Light-year Wavelengths: Insights from Pulsar Timing Arrays}

\author*[1]{\fnm{Stephen~R.} \sur{Taylor}}\email{stephen.r.taylor@vanderbilt.edu}

\affil*[1]{\orgdiv{Department of Physics \& Astronomy}, \orgname{Vanderbilt University}, \orgaddress{\street{2301 Vanderbilt Place}, \city{Nashville}, \postcode{37235}, \state{TN}, \country{USA}}}

\abstract{Arrays of precisely-timed millisecond pulsars are used to search for gravitational waves with periods of months to decades. Gravitational waves affect the path of radio pulses propagating from a pulsar to Earth, causing the arrival times of those pulses to deviate from expectations based on the physical characteristics of the pulsar system. By correlating these timing residuals in a pulsar timing array (PTA), one can search for a statistically isotropic background of gravitational waves by revealing evidence for a distinctive pattern predicted by General Relativity, known as the Hellings \& Downs curve. On June 29 2023, five regional PTA collaborations announced the first evidence for GWs at light-year wavelengths, predicated on support for this correlation pattern with statistical significances ranging from $\sim\!2-4\sigma$. The amplitude and shape of the recovered GW spectrum has also allowed many investigations of the expected source characteristics, ranging from a cosmic population of supermassive binary black holes to numerous processes in the early Universe. In the future, we expect to resolve signals from individual binary systems of supermassive black holes, and probe fundamental assumptions about the background, including its polarization, anisotropy, Gaussianity, and stationarity, all of which will aid efforts to discriminate its origin. In tandem with new facilities like DSA-2000 and the SKA, fueling further observations by regional PTAs and the International Pulsar Timing Array, PTAs have extraordinary potential to be engines of nanohertz GW discovery.}




\maketitle

\setcounter{tocdepth}{2}          
\renewcommand{\contentsname}{Contents} 
\begingroup
  \hypersetup{linkcolor=blue}    
  \tableofcontents
\endgroup
\bigskip

\section{Introduction}\label{sec1}

The pace of gravitational wave (GW) discovery has quickened in the decade since the first detection in 2015 \citep{PhysRevLett.116.061102}. This was almost a century since their prediction by Einstein \citep{einstein1916naherungsweise} as a consequence of his general theory of relativity (GR) that enshrines gravity within a geometric description of spacetime \citep{einstein1915feldgleichungen}. Yet the intervening century was marked by necessary theoretical and technical developments \citep[e.g.,][]{kennefick2016traveling}, such as accurate waveform models, detector design and refinement, as well as indirect evidence through the observed orbital decay of a binary pulsar system in extraordinary agreement with GW emission predictions. It seems historically fitting in light of this review that pulsar astronomy provided not only the first evidence for GW emission but also became key to GW detection.

The production of GWs requires a source with a non-zero accelerating quadrupole moment of its mass-energy distribution. As such, GWs are not emitted from radially pulsating or linearly moving stars. Yet two stars orbiting one another possess the required time-varying quadrupole mass moment, and are in fact the only unambiguous physical sources currently detected. Nevertheless, it is only binary compact objects that are within the sensitivity of current detectors. This can be understood by how the GW strain amplitude scales with the orbital characteristics of the binary, which increases with the mass of the constituents and as the orbital period decreases. Regular stars simply cannot get close enough to each other to produce detectable GWs without interacting, whereas systems of objects like white dwarfs, neutron stars, and black holes have far greater compactness. 

Gravitational waves have many properties of more quotidian waves; they have amplitudes, frequencies, and polarization. The restriction of gauge freedoms in GR allows GWs to have only two polarizations, referred to as the $+$ and $\times$ modes. These produce spatial distortions in the plane orthogonal to the direction of wave propagation, and whose patterns of distortion are offset by $\pi/4$ from one another. There are clear analogies with electromagnetic waves, which are also transverse waves propagating at $c$.

One can journey through the GW spectrum guided by the mass of the sources. The innermost stable circular orbit reached by a binary system, after which the system would plunge quickly toward coalescence, has objects separated in proportion to the binary's total mass. At mass scales of $\sim\!1-200M_\odot$ the frequency of GW emission close to coalescence aligns with the sensitivity of ground-based laser interferometers such as LIGO\footnote{\href{https://www.ligo.caltech.edu/}{https://www.ligo.caltech.edu/}}, Virgo\footnote{\href{https://www.virgo-gw.eu/}{https://www.virgo-gw.eu/}}, and KAGRA\footnote{\href{https://gwcenter.icrr.u-tokyo.ac.jp/en/}{https://gwcenter.icrr.u-tokyo.ac.jp/en/}} (known collectively as the LVK Collaboration). The first detection of GWs in 2015 \citep{PhysRevLett.116.061102} was of the inspiral and coalescence of two black holes with masses of $\sim\!36M_\odot$ and $\sim\!29M_\odot$. Since that time, there have been $\sim300$ confirmed detections\footnote{\href{https://gwosc.org/eventapi/html/GWTC/}{https://gwosc.org/eventapi/html/GWTC/}} of (mostly) binary black hole mergers, along with some neutron star-black hole and binary neutron star mergers, including a dramatic multi-messenger binary neutron star coalescence producing many electromagnetic signatures in tandem with the measured GW signal \citep{PhysRevLett.119.161101}. 

At higher masses, one must move to detectors that have lower frequency sensitivity, like the space-borne Laser Interferometer Space Antenna (LISA) that is scheduled for launch in mid-2035 \citep{2023LRR....26....2A,2024arXiv240207571C}. While some of the more massive stellar-origin binary black-hole systems could be detectable in LISA \citep[e.g.,][]{2025JCAP...01..084B}, a key target will be binary systems of $\sim\!10^4-10^7 M_\odot$ black holes that exist at the centers of most galaxies \citep{2021FrASS...8....7S,2023LRR....26....2A}. However, the most prodigious sources for LISA are expected to be the entire Galactic population of tight-separation binary white dwarf systems \citep{2001A&A...375..890N}. While approximately $10^4$ such systems will be individually resolvable, the majority of signals will sum incoherently below the resolution limit, producing a stochastic GW foreground whose spatial distribution tracks the shape of the Milky Way \citep{2014PhRvD..89b2001A,PhysRevD.101.123021}.

Binary systems of supermassive black holes (SMBHBs) with masses $\sim\!10^8-10^{10}M_\odot$ will merge below the range of LISA sensitivity or with insufficient signal-to-noise ratio. Detecting those requires us to move even lower to the $\sim1-100$~nHz band, where the timing properties of arrays of Milky Way pulsars are searched for correlated influences indicative of low-frequency GWs. This review focuses on pulsar timing arrays (PTAs; \cite[e.g.,][]{taylor2021nanohertz,2018CQGra..35m3001V,2024ResPh..6107719V,2021hgwa.bookE...4V,2025arXiv250500797K}) and their recent breakthroughs. At these low frequencies, GW emission causes the binaries to evolve slowly enough that the signal remains quasi-monochromatic. A large number of such signals produce a stochastic GW background signal whose statistical properties encode the demographics, dynamics, and spatial distribution of the population. In fact, in a simple model of a population of circular binaries evolving purely by GW emission, the ensemble average GW characteristic strain spectrum scales with frequency as $f^{-2/3}$ \citep{2001astro.ph..8028P}. This signal manifests itself in the timing of pulsars as a noise-like process, but, crucially, one that is distinctively correlated between different pulsars.

It is this kind of stochastic signal for which PTA collaborations around the world have recently found compelling evidence, unveiling new terrain in the GW spectrum that is inaccessible to ground-based or even space-borne detectors. Beyond statistical probes of the SMBHB population, this field holds the promise of individually resolvable SMBHB sources with potential electromagnetic-counterpart signatures, and even GW signals from the early Universe, or which probe modifications to GR. In this review, I will give an overview of PTAs and how they work, a brief history of the field leading to the breakthrough results of $2023$, what this means for our knowledge of SMBHB systems and potential cosmological sources, and the exciting scientific questions and facility developments that are driving efforts into the future.    

\section{Pulsar timing arrays}\label{sec2}

Pulsars are a special class of rapidly rotating neutron star, the last stage of compact matter in the Universe before collapsing into a black hole. Neutron stars have radii between $\sim\!10-15$~km and masses approximately between $\sim\!1-2M_\odot$, although the maximum mass is not yet firmly established by observations, and is strongly linked to the ongoing research into the neutron star equation of state \citep[e.g.,][and references therein]{2021PrPNP.12003879B}. The identity of pulsars as rotating neutron stars was unknown at the time of their discovery in 1967 by Jocelyn Bell Burnell as part of a team led by Anthony Hewish \citep{1969Natur.224..472H}. The observations consisted of repeating radio pulses separated by $1.34$~s, while the small parallax and pulse duration implied a compact source at an astronomical distance. It was quickly established that only the ``lighthouse model'' \citep{1968Natur.218..731G} would fit, wherein a rotating neutron star with a misaligned magnetic field---along which particles are accelerated---produces electromagnetic emission, which flashes across our line of sight every rotational period. 

Pulsar astronomy is an incredibly important field in its own right \citep[e.g.,][]{2004hpa..book.....L,2022ARA&A..60..495P,2024LRR....27....5F}, and has been enormously influential in attempts to measure the neutron star equation of state, understand the ionized interstellar medium, probe the interaction of electromagnetic fields under extreme gravity conditions, and (most pertinent to this review) measure the orbital evolution of binary systems emitting GWs. In $1974$, Russell Hulse and Joseph Taylor used the Arecibo radio telescope to discover a pulsar in a binary system with another companion neutron star \citep{1975ApJ...195L..51H}. Repeated pulsar timing observations allowed the characteristics of this system to be reconstructed, displaying a $7.75$~hour orbit experiencing a shift in periastron time that was in excellent agreement with calculations based on the system's production and emission of GWs \citep{1982ApJ...253..908T}. This extraordinary binary pulsar discovery, combined with the long-term observing campaign, provided the first unambiguous evidence for GWs, even if it was evidence of emission rather than reception.

Shortly after the discovery of the Hulse-Taylor binary, other researchers were exploring pathways for GW detection that were alternatives to the emerging picture of ground-based laser interferometry (as first discussed in Rai Weiss' 1972 MIT technical report; \cite{weiss1972quarterly}). Notably, \citet{1975GReGr...6..439E} calculated how GWs passing between Earth and a distant spacecraft would affect the propagation of electromagnetic signals between them, laying the groundwork for considerations of precise timing measurements for GW detection. Subsequently, \citet{1978SvA....22...36S} proposed how pulsar timing measurements could be used to detect the influence of long-wavelength GWs from an intervening binary system, while \citet{1979ApJ...234.1100D} generalized the concept to consider the timing influence of GWs from more distant sources incident upon an Earth-pulsar system. The latter is the framework we now use. Yet, the major impediment was the limited timing accuracy of pulsars, which would be necessary to search for weak-amplitude GW signals.

In 1982 a new class of pulsar was discovered by a team led by Don Backer \citep{1982Natur.300..615B,2024JAHH...27..465D}, dubbed millisecond pulsars (MSPs) for their much faster rotation than canonical pulsars. Interestingly, Ron Hellings and George Downs did not seem to know about this discovery at the time they proposed their 1983 method of cross-correlating pulsar timing residuals to compare with the expected signature of a statistically isotropic background of GWs, now known as the Hellings \& Downs curve \citep{1983ApJ...265L..39H}. 

MSPs offer greater timing precision, fewer glitches, and lower spin instability, making them far more suited to precision timing campaigns for GW searches. By the end of the decade, this prospect was being explored, first proposed by \citet{1989ASIC..262..113R}, then studied in detail and formalized by \citet{1990ApJ...361..300F}. From 1990 onward, MSPs were used to place ever more constraining limits on the cosmic background of GWs \citep{1990PhRvL..65..285S,1994ApJ...428..713K,1996PhRvD..53.3468T,1996PhRvD..54.5993M,2002nsps.conf..114L,2004ApJ...606..799J,2006ApJ...653.1571J}; in the mid-2000s, these efforts began to be formalized within regional PTA collaborations. Between 2005 and 2007, the Parkes Pulsar Timing Array (PPTA; Australia; \cite{2013PASA...30...17M}), the European Pulsar Timing Array (EPTA; \cite{2016MNRAS.458.3341D}), and the North American Nanohertz Observatory for Gravitational Waves (NANOGrav; \cite{2009arXiv0909.1058J}) were established. In 2008, the first ``Worldwide Pulsar Timing Array'' meeting took place at Arecibo, later becoming known as the International Pulsar Timing Array (IPTA) upon its establishment in 2009 \citep{2016MNRAS.458.1267V}.

\section{Key concepts} \label{sec3}

\begin{figure}[!t]
    \centering
    \includegraphics[width=0.9\columnwidth]{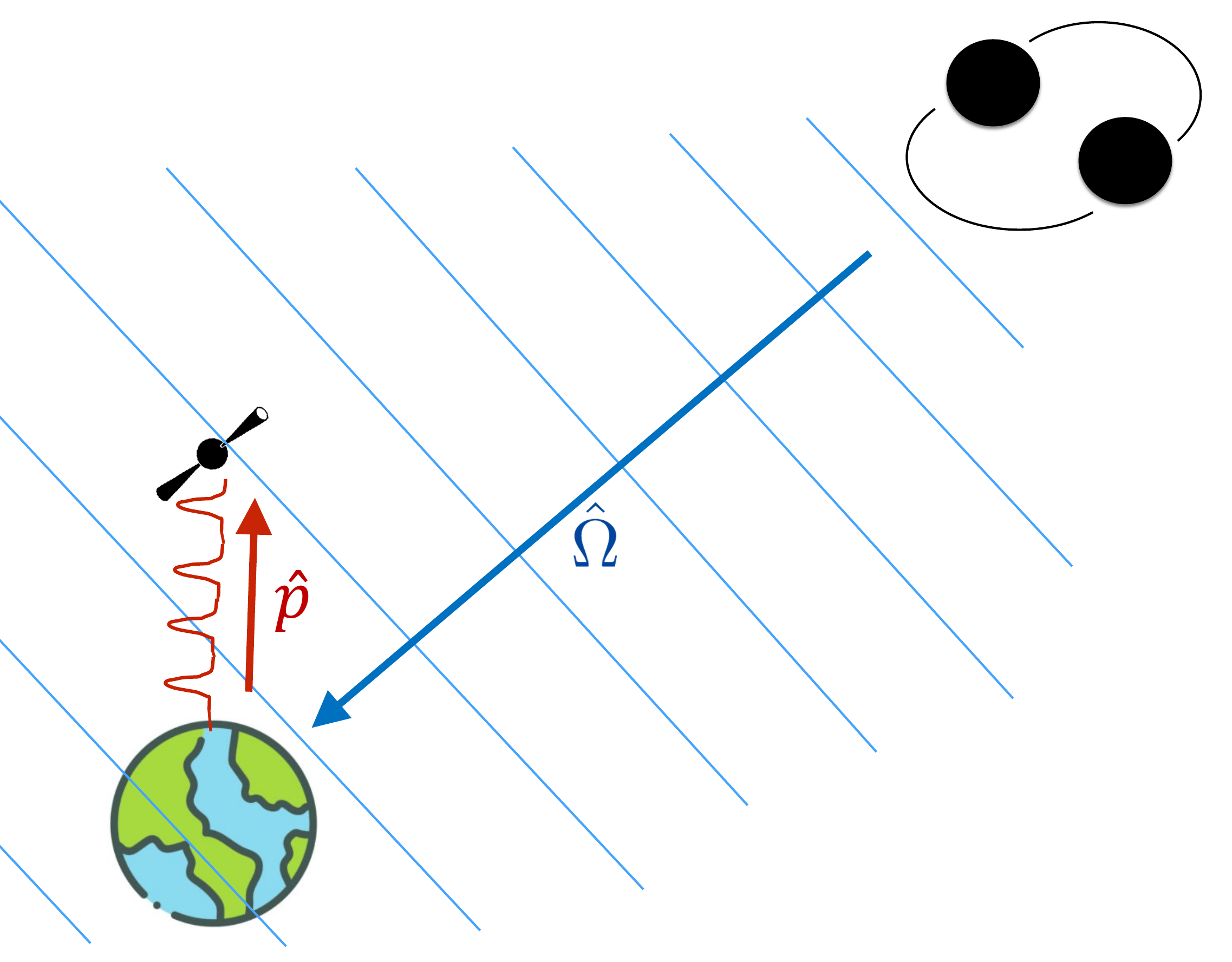}
    \caption{A gravitational-wave of extragalactic origin---here shown coming from a black-hole binary---enters the Milky Way and passes through an Earth-pulsar system. The train of radio pulses that we measure from the pulsar, and which propagate through the perturbed spacetime represented by the GW wavefronts in blue, are represented in red. With Earth (or the Solar System barycenter) as the origin, the pulsar is at distance $L$ in direction $\hat{p}$, while the GW with wavelength $\lambda$ originates at distance $d$ from direction $-\hat\Omega$. The hierarchy of distances is such that $\lambda\ll L\ll d$, where $\lambda$ is of order lightyears (or a parsec), $L$ of order $10^2-10^3$ parsecs, and $d$ is likely greater than a Megaparsec.}
    \label{fig:pta_geometry}
\end{figure}

It is necessary to set the stage with some preliminary equations and concepts, summarized and adapted from \citet{maggiore2008gravitational}, \citet{maggiore2018}, and \citet{taylor2021nanohertz}. We adopt the following line element for perturbed Minkowski spacetime, where a GW is described in transverse traceless (TT) gauge affecting spatial dimensions indexed by Roman indices:
\begin{equation}
    ds^{2} = -dt^{2} + \bigl[\delta_{ij} + h^{\mathrm{TT}}_{ij}(t,\vec{x})\bigr]\,dx^{i}dx^{j}.
\end{equation}

Now, let us imagine photons propagating outward at time $t_\mathrm{em}$, from a pulsar to Earth along a path $-L\hat{p}$, where $L$ is the distance to the pulsar, and $\hat{p}$ is a unit vector pointing from Earth to the pulsar. A diagram showing the setup of the problem is shown in \autoref{fig:pta_geometry}. 

The apparent change in the pulse period, $P$, due to GW perturbations is
\begin{align}
    \Delta P = \frac{\hat{p}^i \hat{p}^j}{2} 
    \int_{t_{\mathrm{em}}}^{t_{\mathrm{em}}+L} dt' 
    &\left\{ h^{\mathrm{TT}}_{ij}\!\left[t' + T, \vec{x}_0(t')\right] 
    - \right.\nonumber\\
    &\quad\left.h^{\mathrm{TT}}_{ij}\!\left[t', \vec{x}_0(t')\right] \right\},
\end{align}
where $\vec{x}_0(t') = \bigl(t_{\mathrm{em}} + L - t'\bigr)\,\hat{p}$. Given that pulsar rotational periods are of order milliseconds, while the GW periods of interest here are years to decades, one can expand the first metric perturbation term to linear order, such that
\begin{equation}
    \frac{\Delta P}{P} = \frac{\hat{p}^i \hat{p}^j}{2} 
    \int_{t_{\mathrm{em}}}^{t_{\mathrm{em}}+L} dt' 
    \left[ \frac{\partial}{\partial t'} h^{\mathrm{TT}}_{ij}(t', \vec{x}) 
    \right]_{\vec{x} = \vec{x}_0(t')}.
\end{equation}

Finally, upon identifying $\Delta P/P\equiv z$ as the Doppler shift to the arrival rate of pulses, one can use an ansatz form of $h^{\mathrm{TT}}_{ij}$ to identify the following general form:
\begin{align}
    z(t, \hat{\Omega}) = \frac{\hat{p}^i \hat{p}^j}{2 ( 1 + \hat{\Omega} \cdot \hat{p} )} \, &\left[ h_{ij}(t, \vec{x}_{\mathrm{earth}}) 
    - \right. \nonumber\\
    & \quad\left. h_{ij}(t - L, \vec{x}_{\mathrm{pulsar}}) \right],
\end{align}
where $\hat\Omega$ is the unit vector direction of GW propagation. The two terms in this expression are often denoted the ``Earth term'' and the ``pulsar term'' respectively. The Earth term's GW phase is common to all pulsars, albeit the term's amplitude varies according to directional response factors encapsulated by the pre-factor to the squared bracket. However, the pulsar term has a GW phase and frequency that depend on the time in the past at which the GW swept past a given pulsar. The lag time in the snapshot of a source's dynamic state represented by the pulsar and Earth terms is $L(1-\cos\mu)$, where $\mu$ is the angular separation between the GW's origin and the pulsar sky location. 

\begin{figure*}[!t]
    \centering
    \includegraphics[width=0.4\textwidth]{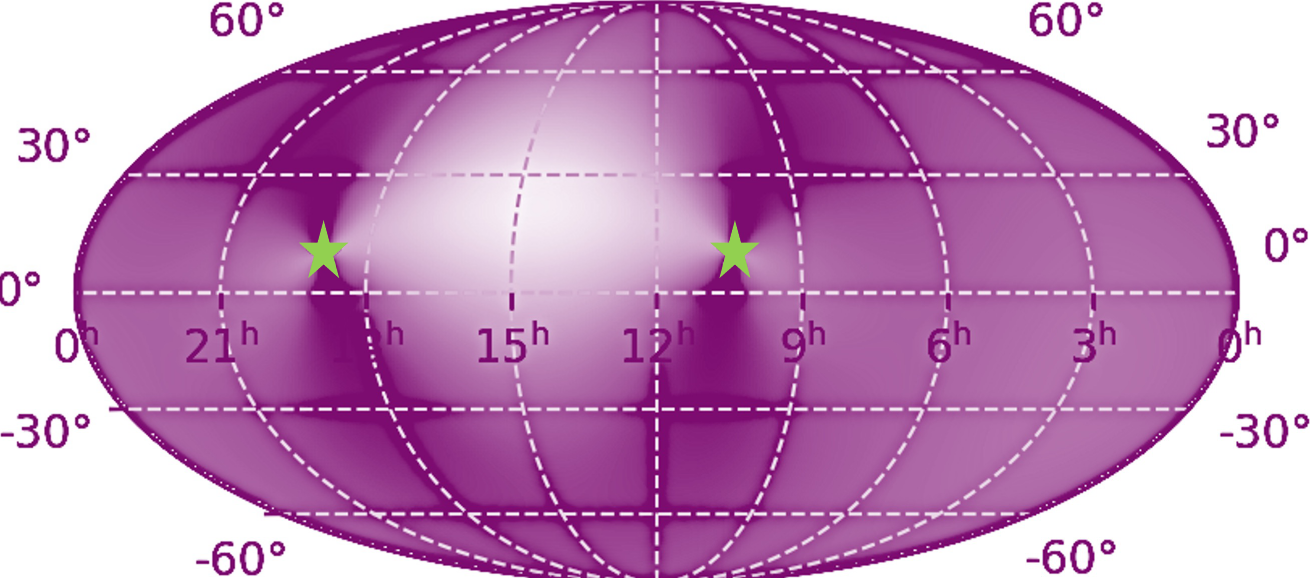}\hspace{10pt}
    \includegraphics[width=0.4\textwidth]{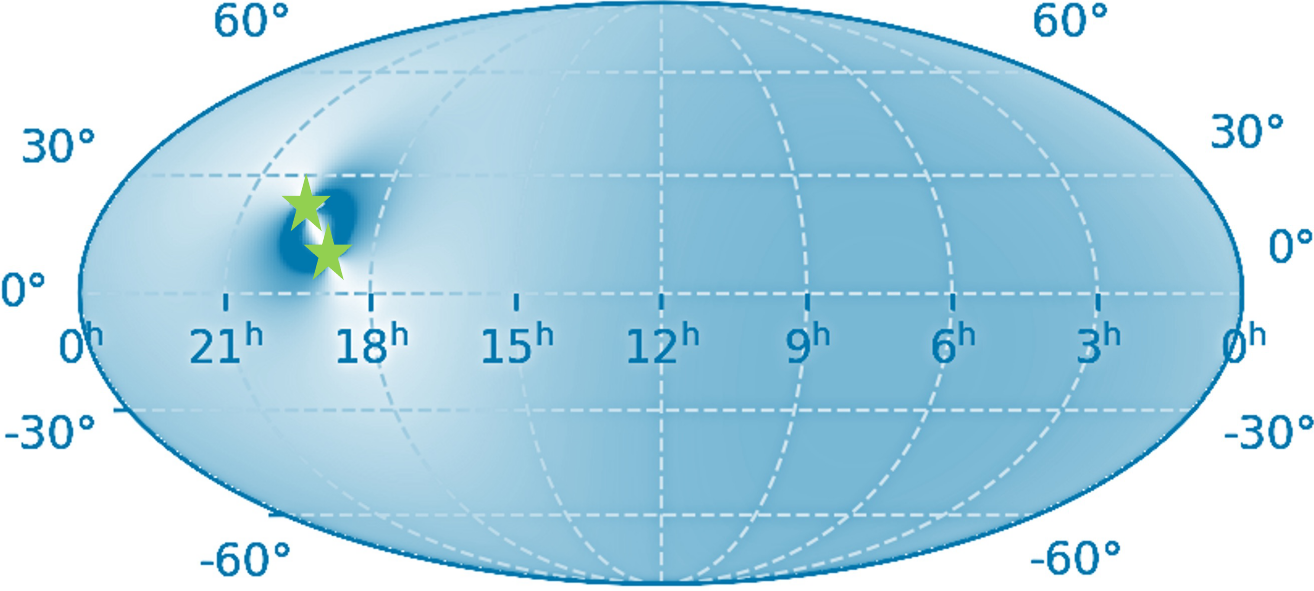}\vspace{10pt}\\
    \includegraphics[width=0.4\textwidth]{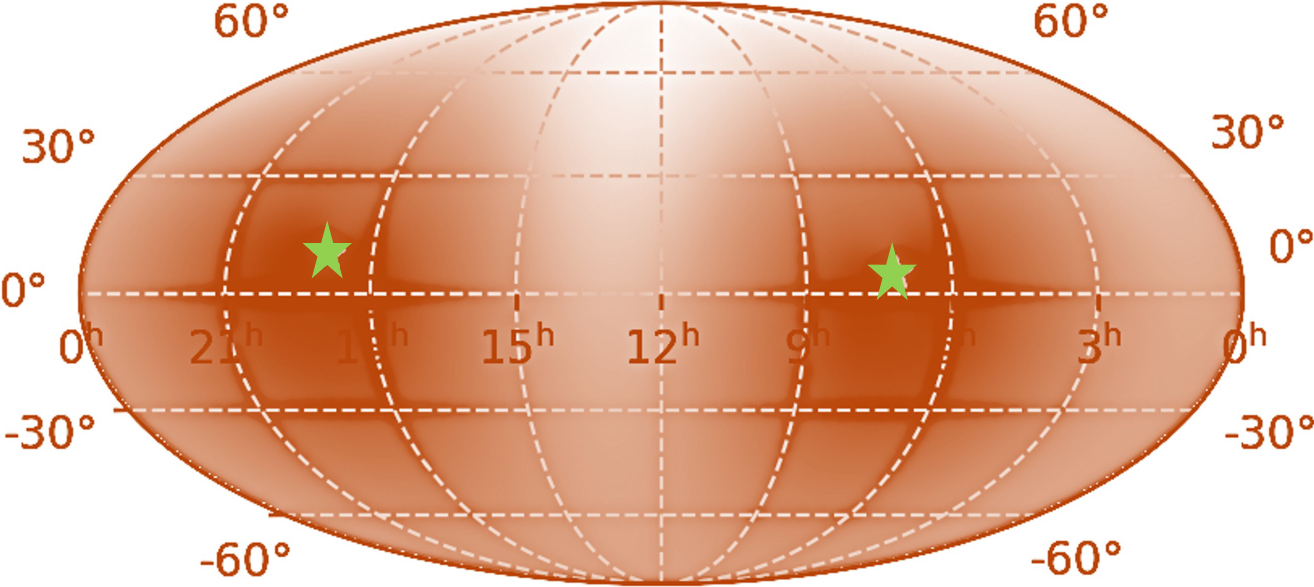}\hspace{10pt}
    \includegraphics[width=0.4\textwidth]{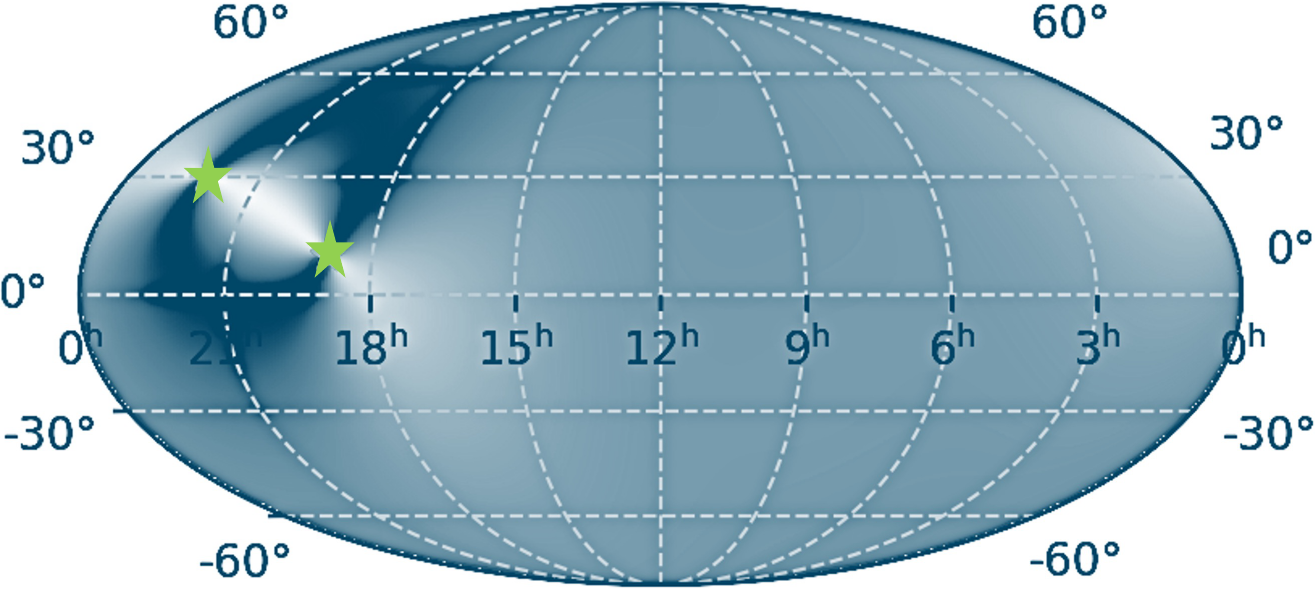}\vspace{10pt}\\
    \hspace{-21pt}\includegraphics[width=0.4\textwidth]{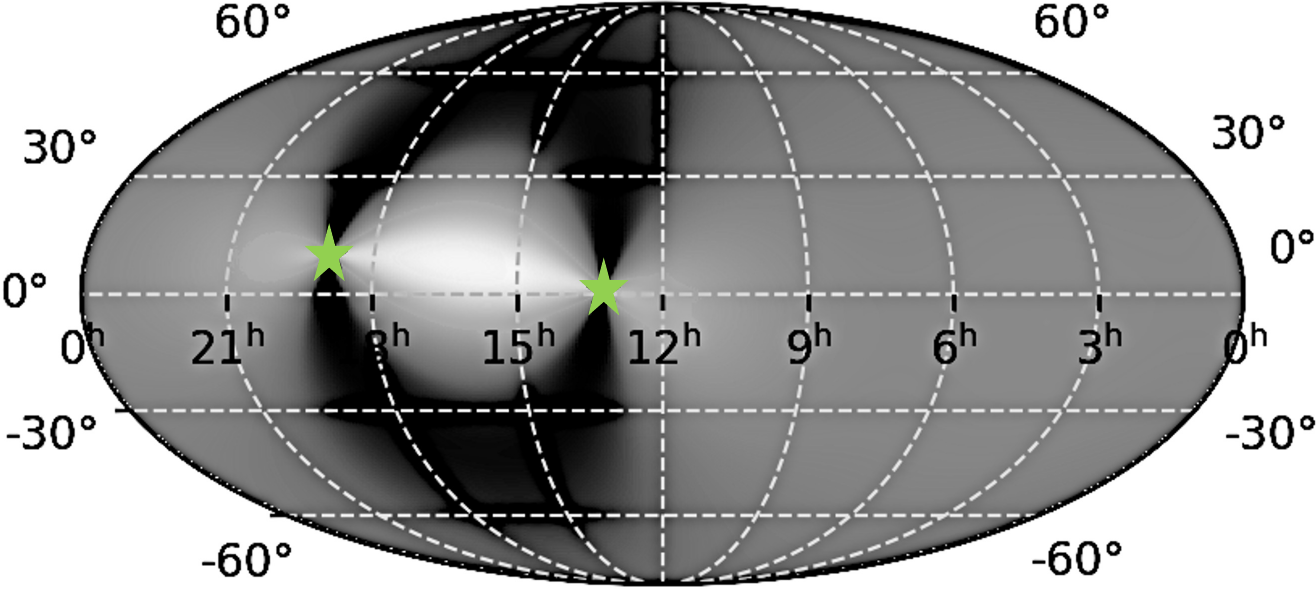}\hspace{40pt}
    \includegraphics[height=70pt,width=0.25\textwidth]{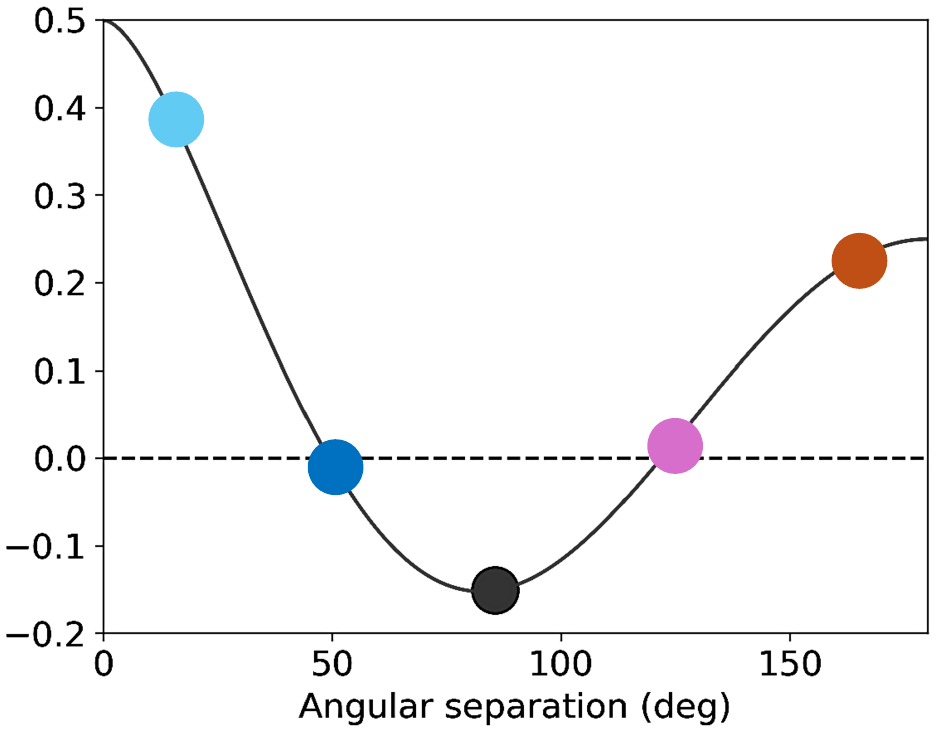}
    \caption{A visual primer on the Hellings \& Downs curve. For a given pair of pulsars (green stars), the Hellings \& Downs coefficient is given by the correlated response functions of the pulsars, averaged over the (isotropic) distribution of GWB power on the sky. Each panel showing a map displays the correlated responses of a pair of pulsars, which when integrated over the sky, gives the Hellings \& Downs coefficient with the matching color in the bottom right panel.}
    \label{fig:hd_cartoon}
\end{figure*}

One can understand this lag geometrically, using \autoref{fig:hd_cartoon} as a guide. Imagine we track a particular GW wavefront, which passes the pulsar at time $t_p$ just as a radio pulse is sent out from the pulsar toward Earth. The difference in source dynamical states represented by the pulsar- and Earth-term is equivalent to how far that GW wavefront is able to propagate past the Earth in the time, $L$, it takes for the radio pulse to reach Earth at time $t_e$. If the angle between the pulsar and GW source were $\pi/2$, this distance is just $L$. In other words, by the time the pulse reaches Earth, new GW wavefronts are arriving at Earth that were sent out from the source a time $L$ later than when the original  wavefront passed by the pulsar (which has already propagated a distance $L$ beyond Earth). However, if the angle between the pulsar and GW source is some arbitrary $\mu$, we have to subtract the projected distance of the pulsar vector along the line-of-sight to the GW source, which is $L\cos\mu$. Hence, the extra distance that a wavefront travels beyond Earth during the pulse travel time, and thus the lag between wavefronts that interact with the Earth-pulsar system between the emission and reception of the pulse, is $L(1-\cos\mu) = L(1+\hat\Omega\cdot\hat{p})$. Finally, $t_p = t_e-L(1+\hat\Omega\cdot\hat{p})$.

Any GW signal can be decomposed into its polarization basis tensors, $e^A_{ij}$, with associated polarization amplitudes, $h_A$, where $A\in[+,\times]$, such that we can define the Earth-pulsar antenna response function as
\begin{equation}
    F^A(\hat{p},\hat{\Omega}) = \frac{\hat{p}^i \hat{p}^j}{2 ( 1 + \hat{\Omega} \cdot \hat{p} )} e^A_{ij}(\hat\Omega).
\end{equation}

The TOA perturbation is just given by the time integral of the shift to the pulse arrival rate, i.e., $r(t) = \int_0^t dt'z(t')$. When we search for individual GW sources, we adopt a deterministic model for $r(t)$. However, when searching for a stochastic GWB, we instead model it through the two-point correlation statistics of different $r(t)$ across the pulsar array. The ensemble average cross-correlation of the timing perturbations between pulsars $a$ and $b$ can be written as 
\begin{equation} \label{eq:timing_covariance}
    \langle r_a(t) r_b(t') \rangle = \Gamma_{ab}C(|t-t'|),
\end{equation}
where $C(|t-t'|)$ is the covariance in timing perturbations induced by the GWB, which one can relate to a power spectral density, and ultimately to its characteristic strain spectrum, through
\begin{equation}
    C(\tau) = \int_0^\infty df\,S(f)\cos(2\pi f\tau),
\end{equation}
and $S(f) = h_c^2(f) / 12\pi^2f^3$. The other factor in \autoref{eq:timing_covariance} is $\Gamma_{ab}$, related to the overlap reduction function for PTAs. It is essentially the overlap in response functions between different pulsars, averaged over the GWB power distribution on the sky, $P(\hat\Omega)$:
\begin{equation}
    \Gamma_{ab} 
    = \frac{3}{2}(1+\delta_{ab})\int_{S^2} \!\!d^2\hat\Omega\,\, \frac{P(\hat\Omega)}{4\pi} \!\!\!
   \sum_{A=+,\times}\!\!\! 
   F_a^A(\hat\Omega) F_b^A(\hat\Omega).
\end{equation}

The factor of $(1+\delta_{ab})$ is a good approximation to the decorrelation of pulsar terms when pulsars are physically separated by more than $\sim$a GW wavelength. The pre-factor of $3/2$ is convention, such that $\Gamma_{aa}=1$. Assuming a statistically isotropic GWB with $P(\hat\Omega)=1\,\,\,\forall\,\hat\Omega$, $\Gamma_{ab}$ becomes the well-known Hellings \& Downs curve,
\begin{equation}
    \Gamma_{ab} = \frac{3}{2} x_{ab} \ln(x_{ab})
   - \frac{1}{4} x_{ij}
   + \frac{1}{2}
   + \frac{1}{2} \delta_{ab},
\end{equation}
where $x_{ab} = (1-\cos\mu_{ab})/2$. The Hellings \& Downs curve plays a central role in this review, and indeed the field, as it is the distinctive signature of a GWB in PTA data. Spectral signatures can be mimicked, but this spatial signature less so \citep{2016MNRAS.455.4339T}. Some intuition for the meaning of the Hellings \& Downs curve can be gained from \autoref{fig:hd_cartoon}--- it is the result of observing a statistically isotropic GWB signal through the correlated response functions of pairs of pulsars.

\section{The path to discovery}\label{sec4}

\subsection{Upper limits}

Until 2020, PTAs reported upper limits on the strain amplitude of a putative GWB in the absence of any measured common stochastic process (whether inter-pulsar correlated or not). These limits are usually reported for a GWB signal with a power-law characteristic strain spectrum $h_c(f)=A(f/1\,\mathrm{yr}^{-1})^\alpha$ whose index is fixed to the ensemble average for a circular GW-driven SMBHB population, $\alpha=-2/3$, and whose amplitude is referenced to a frequency of $1\,\mathrm{yr}^{-1}$. 

In 2006, the PPTA used seven MSPs (most of which were timed by the Murriyang telescope at Parkes observatory, but with some public data taken by the Arecibo telescope) to derive a $95\%$ upper limit on the amplitude of the GWB of $9\times10^{-15}$ \citep{2006ApJ...653.1571J}. But it was in the 2010s that the field of PTAs really found its footing, experiencing a decade of competitive cooperation between the regional PTAs in placing ever more stringent upper limits on the GWB.

The EPTA's first limit was published in 2011, using five MSPs timed over $\sim\!~10$~years to place a $95\%$ constraint of $A=6\times10^{-15}$ \citep{2011MNRAS.414.3117V}. This study set the template for how analyses are performed today, constructing a Bayesian model of the pulse times of arrival, analytically marginalizing over timing model parameters, and searching for hyperparameters of the GWB simultaneously with intrinsic pulsar spin noise and other noise effects. This was followed by a limit of $3\times10^{-15}$ in 2015 using $18$~years of observations from six MSPs \citep{2015MNRAS.453.2576L}. However, in retrospect, the posterior distribution for the GWB spectral parameters seems constrained enough to constitute a measurement---at least of a common uncorrelated stochastic process---rather than an upper limit. This would become important several years later.

NANOGrav's first published dataset in 2013 was $\sim\!5$~years in duration, using $17$ MSPS to place a $95\%$ upper limit of $A=7\times10^{-15}$ \citep{2013ApJ...762...94D}. As NANOGrav grew as a collaboration (first funded by a US National Science Foundation PIRE grant, then as a National Science Foundation Physics Frontier Center), it also expanded its array of MSPs, led by a strategy of forging as many pairs of pulsars as possible to drive the signal-to-noise ratio of the GWB higher \citep{2013CQGra..30v4015S}. In 2015, NANOGrav published a nine-year dataset \citep{2015ApJ...813...65N}, where $37$ MSPs were used to place a limit of $1.5\times10^{-15}$ \citep{2016ApJ...821...13A}. The extension of the data set and the expansion of the array certainly helped in bringing about this more constraining limit, as did the recent development of more powerful Bayesian statistical methodologies. 

This analysis also included a search for potential low-frequency spectral departures from a power law, which could indicate environmental (i.e., non-GW) drivers of SMBHB orbital evolution at the corresponding wider orbital separations. This would be expected for binaries evolving through the final parsec of their evolution, until eventually, GW radiation reaction becomes the dominant influence. However, no evidence was found at that time, and it remains elusive still. In fact, the PPTA's 2015 upper limit of $1\times10^{-15}$ using four pulsars \citep{2015Sci...349.1522S} was reported as evidence of stalling or dramatic spectral attenuation from binary environmental mechanisms, but is now more likely to be an artifact of selection effects, e.g., the choice of Solar System ephemeris, and the small number of pulsars employed, which were chosen to give the tightest upper limits.

Indeed, when NANOGrav published its $11$-year limit in 2018 \citep{2018ApJ...859...47A}, the limited accuracy of Solar System ephemeris models resulted in only a mild update to GWB constraints, corresponding to $1.45\times10^{-15}$ using 45 MSPs. The Solar System ephemeris is a vital component of building a pulsar's timing model, and in tracing each pulse backward from reception at Earth, to the Solar System barycenter, and to the pulsar itself. These ephemeris models are produced quasi-regularly, synthesizing legacy and modern data from spacecraft probes or other sources, and sometimes weighted to satisfy a particular mission requirement. None of these requirements included knowing the absolute position of the Solar System barycenter to within an extraordinary accuracy of $\sim\!100$~m, as needed by PTAs to detect the expected amplitude of a GWB. Systematic differences in the ephemeris models resulted in upper limits that appeared to vary from model to model. Eventually, a perturbative correction model was developed to bridge these systematics\footnote{This description hops over almost a year of the author's research life, working with colleagues at JPL and Caltech led by Michele Vallisneri, to develop a working model.}, involving the masses of the gas giants and Jupiter's orbital elements becoming part of the PTA Bayesian fit \citep{2020ApJ...893..112V}. While this resulted in a modest improvement in limits, the experience also signaled that PTAs were on the right track, improving their sensitivity to the point that expected noise systematics were being encountered. 

In addition to individual regional PTA efforts, in 2016 the International Pulsar Timing Array published its first combined data release \citep{2016MNRAS.458.1267V} of NANOGrav, EPTA, PPTA, and publicly available data from \citet{1994ApJ...428..713K}. This data release of $49$ MSPs consisted of observations ranging from $1986$ to $2013$. However, the observational coverage of individual pulsars was sparser than this stretch of time would seem to imply, with multi-year gaps in several pulsars. To save computational time and avoid noise complexities in the more ill-timed pulsars, this analysis used only the four best-timed pulsars to place a limit of $A<1.7\times10^{-15}$ at $95\%$ credibility.

\subsection{A common signal}

While work was wrapping on the analysis of the NANOGrav $11$-year dataset, explorations were already beginning on its $12.5$-year dataset \citep{2021ApJS..252....4A,2021ApJS..252....5A}.\footnote{These naming conventions for the NANOGrav datasets are only guides, and are usually closer to the maximum baseline of a single pulsar. The total timing baseline of the NANOGrav $11$-year dataset is closer to $11.4$ years, and the $12.5$-year dataset is closer to $12.9$ years. The most NANOGrav $15$-year dataset has a timing baseline of just over $16$ years.} This contained $47$ MSPs, but only $45$ had timing baselines greater than three years, and thus would be influential for constraints below GW frequencies of $10$~nHz. It was very quickly realized that the GWB amplitude posterior was inconsistent with zero, and as such, an upper limit would no longer be an appropriate summary statistic to report. In fact the amplitude was measured---not limited---with a posterior median of $A=1.92\times 10^{-15}$ and $90\%$ credible interval of $1.37-2.67\times10^{-15}$ \citep{2020ApJ...905L..34A}. It was discovered that the presence of a stochastic process with a common spectrum across pulsars in the NANOGrav array was statistically favored with Bayesian odds of $\sim\!10^4$ to 1. By contrast, there appeared to be negligible evidence for Hellings \& Downs inter-pulsar correlations, which, as mentioned earlier, provides the distinctive fingerprint of a GWB signal in PTA data. 

This was initially puzzling: how could there be a spectral signature without a spatial cross-correlation one? An explanation was found theoretically \citep{2021PhRvD.103f3027R} and through simulations \citep{2021ApJ...911L..34P}. PTA datasets have been curated by timing a small number of high-accuracy MSPs, then expanded to include ever more pulsars whose accuracy is not quite so impressive as the first group. This means that spectral constraints derived from the long-baseline, well-timed pulsars are encountered first, followed by the weaker signature of inter-pulsar correlations that provides the convincing evidence of a GWB signal. By contrast, curating an array by timing many pulsars contemporaneously from the outset would potentially find cross-correlation evidence first, to be followed later by informative spectral constraints; examples of this approach will be seen later.

NANOGrav's evidence of a common uncorrelated red-noise (CURN) process was shortly followed by similar evidence delivered by analyses of updated datasets from the PPTA \citep{2021ApJ...917L..19G} and EPTA \citep{2021MNRAS.508.4970C}. Using a dataset with a baseline of $24$ years, the EPTA reported a CURN process with $A = 2.95^{+0.89}_{-0.72} \times 10^{-15}$, while the PPTA's analysis of its second data release, consisting of $26$ pulsars timed over a total baseline of $15$~years, found $A = 2.2^{+0.4}_{-0.3} \times 10^{-15}$. This concordance of results built confidence in the results of each regional PTA's analysis, as did the IPTA's unearthing of a CURN signal with $A=2.8^{+1.2}_{-0.8}\times 10^{-15}$ \citep{2022MNRAS.510.4873A}. This was found in the second official data release of the IPTA, which contained $53$ pulsars that had been combined together from older regional PTA datasets \citep{2019MNRAS.490.4666P}. For example, NANOGrav's contribution was its nine-year rather than its $12.5$-year dataset, owing to the significant time and effort needed to robustly combine all regional PTA observations. But still, no Hellings \& Downs correlations were found in this IPTA analysis. 

\subsection{Evidence of Hellings \& Downs correlations}

We now reach a tumultuous period in world events that formed the backdrop for NANOGrav's first tantalizing signs of GWs. Already in late 2019, members of NANOGrav were exploring a preliminary version of its $45$ pulsar $12.5$-year dataset that had been extended to $\sim\!14$ years. This appeared to show a Bayesian odds ratio favoring inter-pulsar correlations (versus simply a CURN process) of a few tens. While uncalibrated detection statistics have limited value, this was exciting enough to marshal the full collaboration in producing a $15$-year dataset that eventually included $68$ MSPs. 

It was in March-May 2020, as the world was rocked by the onset of the COVID-19 pandemic and quarantines, that NANOGrav produced its first compelling recovery of the Hellings \& Downs curve, with a Bayesian odds ratio of several hundreds. This used a preliminary $15$-year dataset of the same $45$ pulsars from the $12.5$-year and preliminary $14$-year datasets.\footnote{SRT: I can vividly remember being on the sofa of my apartment's living room---which doubled as my home office until I eventually used a storage closet for the same purpose---writing messages on Slack with my partner in co-leading NANOGrav's $15$-year GWB search, Dr. Sarah Vigeland of the University of Wisconsin-Milwaukee. The feeling of producing these detection statistics and reconstructed correlation curves was thrilling; almost a feeling of dissociation as we realized the historic nature of the work to which we'd been privileged to contribute.}

NANOGrav spent the next several months running a litany of tests on this preliminary dataset until we felt confident in bringing the results to the International Pulsar Timing Array. These initial results were first discussed in July, then a full presentation made in September 2020, to the IPTA Steering Committee. At the time this was composed of representatives of legacy PTA collaborations (PPTA, EPTA, NANOGrav) and observers from newer PTA groups (the Indian PTA, the South African PTA, and the Chinese Pulsar Timing Array)\footnote{The Indian Pulsar Timing Array officially joined the IPTA in 2021. The South African PTA is now officially APT, the African Pulsar Timing group, which joined the IPTA in April 2025.}. The response was cautious enthusiasm, launching the IPTA on an almost three-year coordinated search campaign. This involved crafting an agreement for each PTA to conduct independent (but coordinated) analyses, and to follow this with a commitment to combine these regional PTA datasets for a subsequent, full IPTA GW analysis. This was known internally as the \textit{3P+ Agreement}.

The next 2-3 years can be summarized by a montage of hard work. After making a commitment and a plan of action, everyone delved into the business of confirming these initial promising results. The various groups in regional PTAs responsible for creating the timing datasets had Herculean tasks, and worked diligently to ensure that the downstream analyses were conducted on high-quality data with robust timing solutions. The GW data analysts sharpened their computational tools, and developed new techniques to strengthen evidence for GW signals and extract details of the underlying sources. The SMBHB astrophysicists in NANOGrav began to develop an open-source population simulation framework\footnote{\href{https://github.com/nanograv/holodeck}{https://github.com/nanograv/holodeck}}, so that the inferred GWB spectrum could be compared against expectations for various SMBHB dynamical or demographic conditions. Moreover, while PTA collaborations had usually included some constraints in their limit papers on potential primordial GW, cosmic string, or early-Universe phase-transition origins, it was in the period 2021-2023 that members of the high-energy theory and cosmology community began to seriously engage with PTA prospects to probe these signals. These colleagues now form a large part of the NANOGrav membership and have a dedicated working group to explore these possible source scenarios.

While the pace of work was not always frenetic in the lead up to the publication and announcement of the results of this coordinated campaign, the final six months certainly were. The CPTA joined this campaign in Spring 2023, aligning the announcement of their (at first, quite surprising) results with the IPTA's. A detection committee composed of internal IPTA members and external experts made several important suggestions that improved the various PTAs' analyses \citep{2023arXiv230404767A}. This all culminated on 29th June 2023, when NANOGrav \citep{2023ApJ...951L...8A}, the PPTA \citep{2023ApJ...951L...6R}, the EPTA+InPTA \citep{2023A&A...678A..50E}, and the CPTA \citep{2023RAA....23g5024X}, announced their independent evidence for a cosmic background of GWs at light-year wavelengths, as signaled by Hellings \& Downs inter-pulsar correlations. 

\begin{figure*}[!t]
    \centering
    \includegraphics[width=\textwidth]{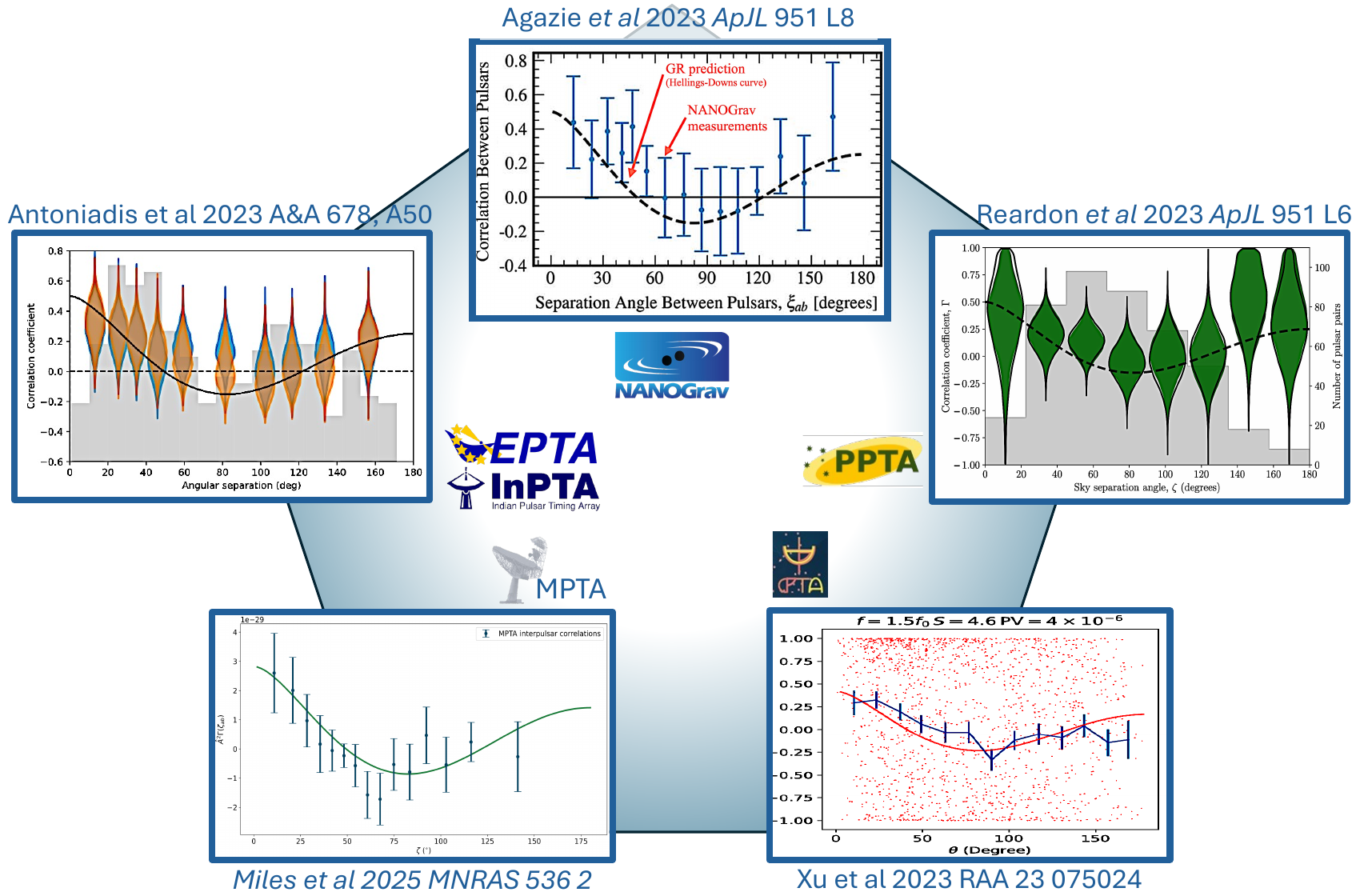}
    \caption{A summary of current evidence for a nanohertz-frequency gravitational-wave background, assessed by consistency of pulsar-pulsar correlations with the Hellings \& Downs curve. Results from NANOGrav \citep{2023ApJ...951L...8A}, the EPTA+InPTA \citep{2023A&A...678A..50E}, PPTA \citep{2023ApJ...951L...6R}, and CPTA \citep{2023RAA....23g5024X} were all released in June 2023, while the MPTA results \citep{2025MNRAS.536.1489M} were released in December 2024.}
    \label{fig:hd_summary}
\end{figure*}

The statistical significance of these correlations varied in tune with the sensitivity of each PTA dataset and the techniques employed. Calibrating the detection statistics and getting $p$-values for cross-correlations in PTA data is challenging, since for a low-frequency GWB there is no time when the signal is absent. Ground-based GW detectors have used the technique of sliding their data to be offset by more than the light-travel time between detectors, then assessing how often spurious noise fluctuations could give the appearance of a coincident detection statistic as large as observed \citep{2010CQGra..27a5005W}. The analogous approaches in PTAs are called sky scrambling \citep{2016PhRvD..93j4047C} and phase shifting \citep{2017PhRvD..95d2002T}, where we pretend pulsars are in different sky locations or have random phase offsets from one another, thereby decorrelating potential inter-pulsar signals. We can also create many synthetic realizations of uncorrelated data, and for our frequentist GWB detection statistic, one can analytically compute the ensemble distribution in the absence of inter-pulsar correlations \citep{2023PhRvD.108j4050H}.\footnote{More recently, simultaneous sky scrambling and phase shifting has been dubbed ``super scrambling'' \citep{2023ApJ...956...14D}, while ``unitary scrambles'' have been developed as a complete generalization of the concept \citep{2025arXiv250610811V}.} 

The results for each PTA are dependent on modeling details and the choice of detection statistic; NANOGrav reported $3\sigma\!-\!4\sigma$, the EPTA+InPTA reported $\sim\!3\sigma$, the PPTA reported $\sim\!2\sigma$, and the CPTA (using a different detection statistic and only deriving evidence from one GW frequency) reported $\sim\!4.6\sigma$ significance.\footnote{CPTA data have only become available to IPTA researchers as of July 2025, and the results have not yet been verified.} The EPTA, InPTA, NANOGrav, and PPTA results were assessed on equal footing in an IPTA analysis \citep{2024ApJ...966..105A}, using comparable noise and signal modeling approaches. Most noise parameters for commonly timed pulsars were consistent, with any tensions likely attributable to different time spans, observational cadence, and radio-frequency coverage. Furthermore, the GWB spectral parameters were found to be within $1\sigma$ agreement using an information tension metric. A pseudo-IPTA analysis (using techniques from \citet{2022PhRvD.105h4049T}) was also performed that effectively extended and supplemented the datasets of each regional PTA with others; this not only resulted in consistent common-process spectral recovery, but also increased the GWB signal-to-noise ratio beyond that of individual regional datasets.\footnote{See also \citet{2025arXiv250320949L} for new IPTA results with pseudo-combination methods.} 

To this list, we add the newer results of the MeerKAT Pulsar Timing Array \citep{2023MNRAS.519.3976M}, which in December $2024$ published their analysis of $83$ pulsars timed over $\sim4.5$~years, recovering $\lesssim3.4\sigma$ \citep{2025MNRAS.536.1489M}. This is a very exciting development, as it shows the power of timing many pulsars from the outset of a PTA campaign, rather than gradually expanding an array that begins from a few great timers. 

For a visual summary of these results, we show the reconstructed inter-pulsar correlation curves from each PTA in \autoref{fig:hd_summary}. These employ either a Bayesian binned parametrization of the correlation signature, or a binned reconstruction of measured pulsar-pulsar pairwise correlations. For example, the NANOGrav binned reconstruction of pairwise correlations gives a $\chi^2$-fit to the Hellings \& Downs curve of $p=0.75$, showing good agreement. This kind of binned reconstruction is a visual indicator of the alignment of results with the expected correlation pattern, but the significance levels reported above use the full information in the PTA datasets rather than just binned values. Moreover these binned visualizations must take into account the presence of the same pulsars in multiple correlation pairs as well as the closeness of some pairs on the sky, which results in covariance of pairwise estimates \citep{2023PhRvD.107d3018A}. This also means that the binned values shown in \autoref{fig:hd_summary} for NANOGrav, MPTA, and CPTA are themselves correlated \citep{2023PhRvD.108d3026A}.  

\section{Interpreting the signal}

What does evidence of this correlation pattern really prove? The Hellings \& Downs curve is the ensemble average pattern of correlations for a statistically isotropic, Gaussian, unpolarized, stationary GW background signal. Noise sources in each pulsar are uncorrelated across the array, while systematics such as clock referencing errors or Solar System ephemeris errors could produce constant or dipolar correlations, respectively \citep{2016MNRAS.455.4339T}. The distinctiveness of the Hellings \& Downs curve, being mostly quadrupolar yet with notable asymmetry, is why it lies at the center of our detection schemes. 

However, deviations from this curve inform departures from its assumptions, e.g., one can search for statistical anisotropy and polarization in the GWB, and reconstruct maps of its intensity \citep[e.g.,][]{2013PhRvD..88f2005M,2013PhRvD..88h4001T,2014PhRvD..90h2001G,2015PhRvL.115d1101T,2020PhRvD.102h4039T}. NANOGrav's recent analysis found no evidence yet for anisotropy, though placed upper limits on the angular power at intensity spherical-harmonic multipoles such that $C_{l>0}/C_{l=0}\lesssim0.2$ \citep{2023ApJ...956L...3A}. Furthermore, deviations from isotropy were approximately two times more constrained near the region of highest pulsar density compared to the opposite side of the sky. One may also expect that the level of intensity anisotropies would increase with GW frequency if the origin of the GWB signal is a population of SMBHBs \citep{2024PhRvD.109l3544S,2024ApJ...965..164G}; the lower occupation levels of higher GW-frequency resolution bins---due to the faster evolution of binaries---are such that the stochasticity of the signal breaks down \citep{2008MNRAS.390..192S}. While PTA datasets are not yet sensitive enough to measure this effect, it could be an important way to discriminate the nature of the GWB's origin. Beyond constructing maps of the GWB intensity as a function of frequency, anisotropy may also herald an individual SMBHB signal on the threshold of detection \citep{SchultEtAl_PTA_FirstBinary_I_inprep,PetrovEtAl_PTA_FirstBinary_II_inprep_2025}. 

In fact, recent theoretical work has examined the competing influences of large-scale structure---the intrinsic clustering of galaxy hosts of SMBHBs---and GW ``shot noise''---directions of high GW luminosity due to the presence of massive or relatively nearby binaries---finding that the latter should be the dominant expected contributor \citep{2025JCAP...03..011G,2024PhRvD.110l3507A}. Therefore the expected GW angular power spectrum in the PTA band is flat and ``white''; this is a consequence of the presence of one or a few bright binaries dominating power maps, and can be derived using the spherical-harmonic addition theorem \citep{2013PhRvD..88h4001T,2024PhRvD.109l3544S}. The PTA community's search for bright single sources is becoming ever more a hot topic as one of the next big milestones for the field, so I will discuss this in detail in \S\ref{sec6}. In the next two subsections I will review the current interpretation of the stochastic GWB signal in terms of its potential origins, which is predominantly performed through its measured spectrum.

\begin{figure*}[!t]
    \centering
    \includegraphics[width=\textwidth]{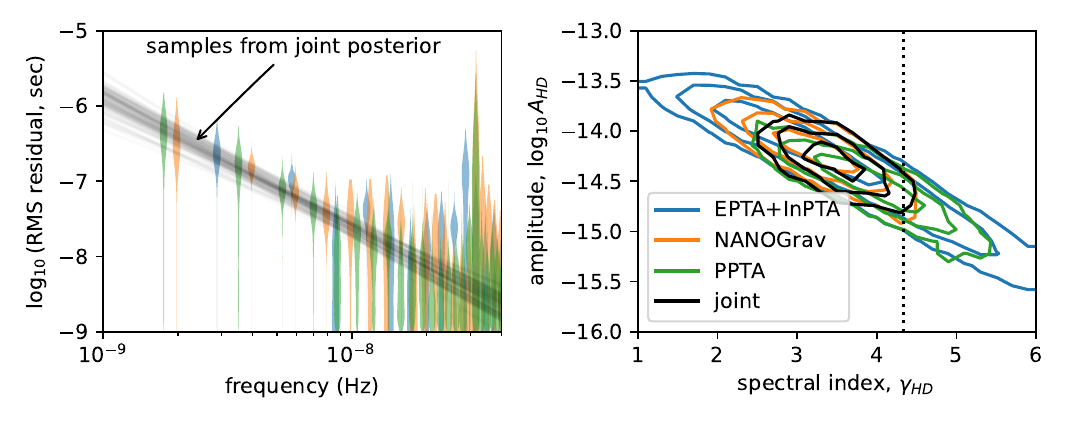}
    \caption{The Bayesian spectral recovery of the GWB signal from the EPTA+InPTA, NANOGrav, and the PPTA. On the left, the power spectral density of induced timing residuals is converted into an RMS timing noise excess, and shown using ``violins'' that represent the posterior spread at each frequency. On the right, these constraints are cast into the parameter space of a power-law spectrum. The dashed vertical then represents the fiducial power-law exponent for the ensemble-averaged power spectral density of timing residuals due to GWs from a population of circular supermassive black-hole binaries, inspiraling due to GW emission. This is $\gamma=13/3$, which is equivalent $h_c\propto f^{-2/3}$. In both panels, a simple product of all PTA constraints approximates the joint spectrum in black dashed on the left, and the black power-law parameters contours on the right. Figure reproduced from \citet{2024ApJ...966..105A}.}
    \label{fig:ipta_compare}
\end{figure*}

\subsection{Characterizing the population of supermassive binary black holes}
The guiding spectral behavior by which PTA collaborations intepret their results is based on the ensemble-averaged expectation for the GW energy density spectrum (as a fraction of closure density), $\Omega_\mathrm{GW}(f)$, from a population of circular inspiraling SMBHBs whose orbital evolution is driven entirely by GW radiation reaction. Several key factors feed into this calculation, which is encapsulated by
\begin{align} \label{eq:omegagw}
    \Omega_\mathrm{GW}(f) &= \frac{1}{\rho_c}\frac{d\rho_{\mathrm{gw}}(f)}{d\ln f}
    = \frac{\pi}{4G\rho_c} f^2 h_c^2(f) \nonumber\\
    &= \int \frac{dz d\mathcal{M}}{1+z} \frac{d^2n}{d\mathcal{M}\,dz}
    \left. \frac{dE_{\mathrm{gw}}(\mathcal{M}, f_r)}{d\ln f_r} \right|_{f_r = f(1+z)}
\end{align}
where $E_\mathrm{gw}$ is the energy emitted in GWs by a single binary, $\rho_c$ is the closure density of the Universe, $f$ is GW frequency (where $f_r$ is rest-frame frequency), $z$ is redshift, $n$ is the number density of SMBHBs, and $\mathcal{M}=(m_1 m_2)^{3/5} / (m_1 + m_2)^{1/5}$ is the chirp mass of those binaries with component masses $m_{1/2}$. The factor $dE_\mathrm{gw}/d\ln f_r$ describes the GW emission from a single binary across logarithmic frequency bins, and is known theoretically from leading-order GW emission calculations as $dE_\mathrm{gw}/d\ln f_r \propto \mathcal{M}^{5/3} f_r^{2/3}$. We then note that frequency scaling behavior can be pulled entirely out of the integral over chirp mass and redshift, allowing one to identify the ubiquitous power-law relationship $h_c(f)\propto f^{-2/3}$.

The power-law strain spectrum searches conducted by all PTA collaborations are broadly consistent with this spectral behavior, as shown in Figure \autoref{fig:ipta_compare}. Yet there are some mild discrepancies in both the spectral index and the amplitude. Most PTA results exhibit a shallower spectrum than expected, but with decent overlap of the posterior distribution with the fiducial value of $-2/3$. NANOGrav's results have the shallowest spectral-index posterior, peaking around $-0.1$ but with $-2/3$ within the $99.7\%$ posterior credible spread. Does this then mean that there is something non-standard about the SMBHB population? Could binaries be driven by their astrophysical environments rather than GW emission at frequencies within the PTA band? Or are SMBHBs not the dominant source of the GWB signal, with cosmological sources being an alternative?

\begin{figure}[!t]
    \centering
    \includegraphics[width=\columnwidth]{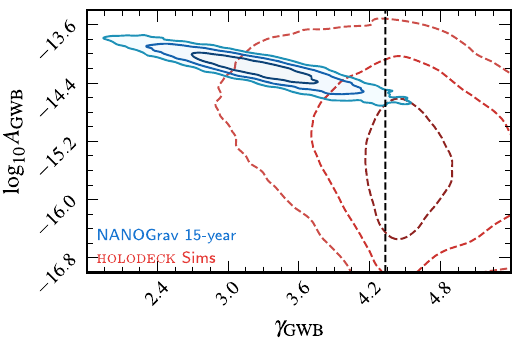}
    \caption{Posterior credible contours from a power-law spectral model of the GWB are contrasted with power-law fits to the strain spectra from many synthesized SMBHB populations, using the \textsc{Holodeck} software package (\href{https://github.com/nanograv/holodeck}{https://github.com/nanograv/holodeck}). The exponent of the ensemble-averaged power-law spectrum is shown as a black dashed line ($\gamma=13/3$). Spectral fluctuations from population discreteness could be a significant mitigating factor when comparing the shallowness of the measured GWB spectrum to the ensemble-averaged spectrum prediction. Figure reproduced from \citet{2023ApJ...951L...8A}.}
    \label{fig:pop_variance}
\end{figure}

Caution must be taken when comparing measured spectra with the fiducial SMBHB-population power-law. Most importantly, this is an ensemble mean prediction; when one accounts for the variance in this prediction, either analytically or through simulations, the distribution of power-law relationships fit to isolated ensembles then overlaps comfortably with all PTA spectral posteriors thus far, as shown in \autoref{fig:pop_variance}. Spectral variance is a topic of much current interest as a means to explain certain low-significance spectral excursions---which may also have a flattening effect on power-law fits---from power-law behavior in NANOGrav's results \citep{2025ApJ...978...31A}. Among other more likely sources (such as improperly modeled noise) these excursions could be viewed as potential unresolved precursor ``bumps'' to resolved binary signals. But more generally, the Poissonian statistics of spectral fluctuations are a probe of the finite nature of the SMBHB population \citep{2003ApJ...583..616J,2025PhRvD.111b3043S,2024ApJ...971L..10L,2025PhRvD.111d3022X}, and can be contrasted against the Gaussian stochastic nature of a cosmologically-sourced GWB signal. The higher moments of the strain power spectral density's ensemble distribution can, like the mean, be similarly expressed as power laws over frequency \citep{2024ApJ...971L..10L}, helping to contextualize future fluctuation studies.

Another mechanism that could appear to flatten a power-law fit to the characteristic strain spectrum is if the ensemble mean spectrum is itself not a simple power-law. As implied above, there is a possibility that SMBHBs could remain coupled to their astrophysical environments at orbital frequencies relevant to the PTA band. The $dE_\mathrm{gw}/d\ln f_r$ term in \autoref{eq:omegagw} can be factored into a dependence on $dt/d\ln f_r$ that encodes the orbital evolution of each system. For binary hardening due to GW emission this scales as $f^{-8/3}$ \citep{peters1964gravitational}, but other mechanisms will create differing behavior. It would be somewhat natural for binary hardening at the low edge of the PTA band to be influenced by non-GW effects \citep[see e.g.,][and references therein]{2019A&ARv..27....5B}; the chain of processes by which galaxies merge, and their resident SMBHs evolve through the tumultuous post-merger environment, involves dynamical friction at large $\sim$~kiloparsec scales, then potentially stellar loss-cone scattering and circumbinary disk torques at sub-parsec separations \citep{1980Natur.287..307B}. 

These latter mechanisms have been invoked to resolve the (misnomered) ``final parsec problem'' \citep{milosavljevic2003final,milosavljevic2003long}. This was originally named due to the theoretical problem of maintaining a supply of stars to the binary scattering loss cone, in order to continue hardening the binary to separations small enough for GW emission to merge it within a Hubble time. Now there are sufficient resolutions to top-up this supply (e.g., triaxiality and rotational stirring of the bulge stellar distribution \cite{2013ApJ...773..100K,2013ApJ...774...87V,2014ApJ...785..163V,2015ApJ...810...49V}) that the problem is more one of using GWs and other observations to determine the precise ingredients and quantities that efficiently merge SMBHs.

If environmental coupling mechanisms are responsible for driving binary hardening at wider separations, then one may expect a positively tilted strain spectrum at low GW frequencies. This would then transition to the usual ensemble-averaged behavior of $f^{-2/3}$ at higher frequencies where GW emission becomes dominant. One can model this semi-agnostically with a turnover model \citep{2015PhRvD..91h4055S} that adds two additional parameters beyond a power law: a parameter to denote the spectral index at lower frequencies, and a parameter that encodes the sharpness of the spectral turnover. However, there is no compelling evidence yet to favor such a turnover model over a simple power-law spectrum, with Bayes factors from NANOGrav's 15-year analysis being $\lesssim2$ \citep{2023ApJ...951L...8A}.

These simple spectral prescriptions conceal many details about the SMBHB population. The turnover model is one way to extract more information about SMBHB binary dynamics, but what about the demographics of the population? To this end, a significant amount of work has been devoted to creating spectral models that express the richness of \autoref{eq:omegagw}, from semi-analytic fitting prescriptions \citep[e.g.,][]{2019MNRAS.488..401C}, to Gaussian-process emulators of SMBHB population simulations \citep{2017PhRvL.118r1102T}, and more recently to dense neural network \citep{2024A&A...687A..42B} and normalizing flow emulators \citep{2025ApJ...982...55L} of simulations. These emulator approaches reproduce not only the ensemble mean spectral shape conditioned on astrophysical hyperparameters, but also the ensemble variance behavior (and the full ensemble distribution in the case of the normalizing-flow approach).

Thus far, the results of directly constraining astrophysical hyperparameters from PTA data are strongly influenced by the GWB's apparent larger-than-expected amplitude. For example, within NANOGrav's $15$-year analysis \citep{2023ApJ...952L..37A}, several hyperparameters that govern the amplitude and efficiency of binary hardening appear to rail against their priors. The GWB amplitude can be explained by higher galaxy merger rate densities ($\sim10^{-4}-10^{-1}$~Mpc$^{-3}$~Gyr$^{-1}$), shorter binary hardening timescales ($\lesssim$~few Gyrs), and higher normalization of the $M_\mathrm{BH}$--$M_\mathrm{bulge}$ relationship\footnote{$\sim$~$10^{8.4} M_\odot$ at a reference stellar bulge mass of $10^{11} M_\odot$, producing a greater abundancy of more massive BHs than previously thought.} than the mean value of their respective prior distributions. The combination of these effects is achievable within the reasonable priors placed upon these hyperparameters, but if it were only one hyperparameter doing so, then it would have to adopt extreme values that would make the result suspect.

The value of the GWB strain spectrum amplitude has received considerable attention. It is above the average of expected values, though within the spread of literature predictions. \citet{2024PhRvD.110f3020S} explored the factors needed to reconcile the GWB amplitude with the current estimate of the black-hole mass density (consistent with the integrated quasar luminosity function through the So\l{}tan argument), finding that the typical SMBH mass was of order a few $10^{10}M_\odot$. This is roughly an order of magnitude larger than naive assumptions based on other estimates of the local SMBH mass function. They assert that simply increasing the merger rate is insufficient. This would extremize what is already known about the typical contributing SMBH population for the PTA GWB, i.e., that it is composed of systems in the extreme high-end tail of the mass distribution. While the amplitude tension explored by \citet{2024PhRvD.110f3020S} is rather mild given the many assumptions and uncertainties in the composite variables, it is notable, provoking further questions and work.

\citet{2024ApJ...971L..29L}'s proposed resolution to this tension arises through improved modeling of the high-mass end of the Galaxy Stellar Mass Function (GSMF). They pieced together a census GSMF relation from \citet{2020ApJ...893..111L} for $M_*\lesssim 10^{11.3}M_\odot$ with that of massive galaxies having $M_*\gtrsim 10^{11.5}M_\odot$ from the volume-limited MASSIVE survey, finding that the local GSMF above $\gtrsim 10^{11.5}M_\odot$ is higher than previous studies. They converted this to a black-hole mass function and a scaling relation for SMBHs and galaxy masses. The resulting abundance of SMBHs with masses above $\sim10^9M_\odot$ is consistent with the number of currently known systems, and their predicted GWB amplitude from SMBHBs is consistent with PTA measurements. 

Whether this mild incongruity will be resolved by purely astrophysical arguments or a moderate revision of the reported GWB amplitude (or a meeting in the middle), remains to be seen. Without tipping too far into editorialization, the author's opinion is that we will witness a modest reduction in the reported GWB amplitude from the PTA community. Not by much; a factor of two would be sufficient to loosen any puzzling tensions. This could come about through improved bespoke pulsar noise modeling (especially for ionized interstellar medium noise processes \cite[e.g.,][]{2024ApJ...972...49L}) and potentially through hierarchical priors on intrinsic low-frequency pulsar noise \citep{2024ApJS..273...23V}. In fact, we may be witnessing some of this effect already from explorations of hierarchical noise priors on EPTA data \citep{2025MNRAS.537.3470G,2024arXiv240903627G}. 

This must all be addressed with great statistical care. Astrophysical ``predictions'' should not be unduly circularly influenced by evolving PTA constraints on the GWB. Likewise, PTA constraints should not be steered to fit predictions, since it is to query these very processes, rather than fit with model predictions, that PTAs carry out their GW searches. Much work is still needed to understand the GWB in light of other astrophysical constraints on the SMBH population. PTAs are uniquely positioned to probe these gargantuan systems lying far into the tail of the cosmic distribution of black holes. 

\subsection{Insights into the early Universe and fundamental physics}\label{sec5}

PTA experiments have long been of interest to the cosmology and fundamental physics communities. Part of this stems from the significant constraining power of pulsars when they are timed as components of binary (or even triple) systems on beyond-GR theories, the strong equivalence principle, and the equation of state of neutron star matter \citep{2024LRR....27....5F,2025LRR....28....3Y,2003LRR.....6....5S,2016ARA&A..54..401O,2021ARNPS..71..433L}. But there is an undeniable growing perception of PTA GW astrophysics as an emerging frontier for ``new physics''\footnote{This shorthand description of any source that is not due to GW emission from a population of SMBHBs will be used throughout.} that is complementary to collider experiments and cosmic surveys. The PTA community itself has often presented upper limits on cosmological origins of the GWB alongside those of the more traditional astrophysical SMBHB population source. 

Yet, in the author's view, interest appeared to drastically increase following the discovery of a common process in PTA data circa 2020. One of the first products of emerging engagement between PTAs and this new community was the 2021 publication of NANOGrav's $12.5$~year limits on cosmological phase transitions \citep{2021PhRvL.127y1302A}. NANOGrav now has a dedicated New Physics working group that was inaugurated in 2023, following several years of growth and the publication of NANOGrav's $15$~year search for signals from new physics \citep{2023ApJ...951L..11A}. This is the second-most cited paper from the suite of NANOGrav $15$~year publications after the principal discovery paper, and the latter has in fact derived most of its citations from the new physics community. Other regional PTA collaborations have seen proportionally similar engagement \citep[e.g.,][]{PhysRevLett.131.171001}. The diversity of new-physics signals being probed and the progress made in this area have already been impressive. The following is merely a broad and necessarily incomplete summary of this progress, heavily based on \citet{2023ApJ...951L..11A}.

There are many cosmological signals that are inconsistent with the level of the observed GWB if one were to assume such signals are the only source. Yet, it stretches credulity to imagine that SMBHBs would never merge in our Universe, even if only via multi-body interactions that overcome some binary stalling at wide separations \citep{2017MNRAS.470.4547D,2016MNRAS.461.4419B,2023PhRvD.108j3034B}. In that case, the level of the astrophysical GWB would be smaller than currently observed, but not zero, leaving room for cosmological signals to play a supporting role. Cosmic inflation predicts a primordial GWB from quantum spacetime fluctuations stretched to macroscopic scales, but with a flat energy density spectrum and with an amplitude constrained by CMB \citep{2021PhRvL.127o1301A,2022PhRvD.105h3524T,2020A&A...641A..10P} and Big Bang Nucleosynthesis (BBN; \cite{2020JCAP...03..010F,2018CQGra..35p3001C}) to be far below reported PTA values \citep{2021NatAs...5.1268M,2024arXiv240915572W}. Hence, a primordial GWB under standard inflationary conditions is not consistent with being a major component of the measured signal. However, other inflationary scenarios can be considered with beyond-standard physics (e.g., spectator fields or a stiff equation-of-state era) that are not yet ruled out by PTA data.

A cosmological phase transition is a very well-studied candidate for an early-Universe source of the GWB \citep[][and references therein]{2023arXiv230502357A}. Phase transitions occur via nucleation of bubbles of the new vaccuum state within the old state. When these bubbles collide, they could source GWs, followed by further GWs from plasma sound waves and turbulence. The resulting GW spectrum is not a power-law but rather peaked at a frequency related to the bubble nucleation rate, the temperature of the Universe at the time of the transition, and the number of degrees of freedom at the time of the transition. To produce GWs at nanohertz frequencies the transition temperature would need to be $\sim$~Mev \citep{2021PhRvL.127y1302A,2023ApJ...951L..11A}, which is relatively late in cosmic history, slightly below the QCD scale, and close to the time of BBN. This seems to necessitate beyond-Standard-Model physics, with potentially a phase transition around the MeV scale in a hidden ``dark'' sector. But even this poses problems by potentially affecting the composition and expansion rate of the Universe around the time of CMB and BBN decoupling, which is already strongly constrained by observations of the latter.

Cosmic strings form another compelling origin for nanohertz-frequency GWs \citep[e.g.,][and references therein]{2018CQGra..35p3001C,sousa2024cosmic}, but are also one of the most strongly disfavored. These are long topological defects formed either during symmetry-breaking cosmological phase transitions or as fundamental string-theory strings. Permeating the Universe, these filaments can intersect and chop off small loops that radiate GWs through oscillations under their own string tension, and other localized features like ``cusps'' and ``kinks''. The amplitude of the observed GWB would imply a string tension $G\mu$~$\sim$~$10^{-10}$--$10^{-11}$ for a stable cosmic string of field-theory origin, overproducing GWs at higher frequencies and resulting in tension with many other experiments \citep{2023ApJ...951L..11A}. The Bayes factors for a GWB produced by stable cosmic strings of field-theory origin are $\sim$~$0.1$--$1$, lower than that of cosmic superstrings ($\sim$~$10$--$100$) or domain walls ($\sim$~$1$--$10$) for which there is greater model flexibility.

PTAs also offer a way to probe various dark matter theories of production and composition, but it should be noted that these influences are not from GWs. In particular, if a component of dark matter in the Milky Way is in the form of ultralight scalar or vector bosons \citep{2000PhRvL..85.1158H,2017PhRvD..95d3541H,2020arXiv200501515F,2021A&ARv..29....7F} with masses $\sim10^{-23}-10^{20}$~eV, then they would produce an oscillating gravitational potential that induces sinusoidal variations in pulsar pulse times of arrival with frequency proportional to the particle mass. While PTA constraints remain above the predicted dark matter abundance level in the Milky Way of $0.4~\mathrm{GeV}\cdot\mathrm{cm}^{-3}$ \citep{2010JCAP...08..004C}, the constraints are comparable to laboratory-based torsion-balance and atomic clock tests \citep[][and references therein]{2023ApJ...951L..11A}. Another signature could be from cold dark matter substructure, i.e., a dark matter clump, or even a primordial black hole (PBH), could produce a gravitational effect on Earth or a pulsar, and therefore perturb pulse arrival times \citep[e.g.,][]{2007MNRAS.382..879S,2007ApJ...659L..33S,2019PhRvD.100b3003D,2021arXiv210405717L}. There is no evidence of such clumps or minihalos yet from PTA data, but this constitutes a novel constraining technique for the future. Finally, the abundance and mass distribution of PBHs would be constrained not only by dark matter substructure tests, but also indirectly by the fact that their formation from the collapse of pronounced primordial density fluctuations may be associated with GW production. These so-called scalar-induced GW models \citep{2021Univ....7..398D} seem to fit the observed GWB spectrum quite well, with Bayes factors $\sim10-100$.

This summary is but a sampling of the different new physics tests hence performed and hereafter possible with PTA data. While a purely cosmological origin of the GWB is unlikely, it is feasible for it to play a role in tandem with an astrophysical GWB from a SMBHB population. Yet so far, the constraints and model selection are based entirely on how well the models fit to the GWB intensity spectrum, and on the employed parameter priors. What is missing is a sense of prior odds for these cosmological models in order to rank their \textit{a priori} credibility. We must also go beyond the GWB intensity spectrum to its polarization structure and tests of its anisotropy. If one were to measure significant anisotropy or non-Gaussianity, then that may be indicative of a dominant astrophysical origin from a finite population of GW-emitting SMBHBs. There remains much information to be mined from the spectral, polarization, and spatial properties of the GWB. The challenge is how to construct a convincing argument out of these pieces of information that would unequivocally yield the dominant origin of the nanohertz-frequency GWB, and ultimately help determine its cosmological-versus-astrophysical composition.

\section{Open questions, and the future}\label{sec6}

We are at the dawn of our exploration of the low-frequency GW spectrum. So soon after the first evidence of tell-tale Hellings \& Downs inter-pulsar correlations in pulsar-timing datasets, there is much yet to understand about this nanohertz-frequency GW background. Beyond the questions and emerging directions that have been discussed earlier in this review, there are still several basic assumptions about the GWB that need a rigorous assessment. 

\subsection{The GW background}

Questions about the Gaussianity of the signal were mentioned earlier in the context of its origin either as a finite astrophysical population or a cosmological process. While there have been significant advances in the theoretical framework of this question, practical data-analysis strategies are needed that can link any non-Gaussianity (either inferred through an astrophysical model \citep{2025PhRvD.111d3022X,2025PhRvD.111b3043S,2025JCAP...01..017B} or agnostically through, e.g., Gaussian mixture modeling \cite{2025arXiv250808365F}) back to the source astrophysics. 

Another basic assumption is on the stationarity of the signal; succinctly, this means that the statistics of the GWB do not depend on the absolute time at which observations are made, but only on relative lags between observations. This translates to independent sampling frequencies, such that each frequency resolution bin will probe a different sample of sources. Although this assumption is mostly on solid footing, we must be cautious, since any long-timescale process sampled within an observation window shorter than its correlation timescale will suffer from frequency covariances (i.e., spectral leakage) \citep{2025arXiv250613866C}. On the source modeling side, one may find non-stationarity emerging from a small number of burst-like or eccentric binaries \citep[e.g.,][]{2025PhRvD.111b3047F,2024PhRvD.109l3010F,2025arXiv250614882E,2023PhRvD.108j2007D}, where GW emission spread over harmonics of the binary orbital frequency will couple different sampling frequencies together.

We also assume that the GWB is statistically unpolarized, regardless of whether it is astrophysical or cosmological in origin \citep[however, see, e.g.,][]{2024PhRvD.110j3505J,2024PhRvD.110d3501A,2020PhRvD.102b3004B,2022PhRvD.106b3004S}. This is a good assumption for most cosmological theories, but when considering an astrophysical population, one must remember that, while a single binary is elliptically polarized, the ensemble average of the population should exhibit no preferred polarization. However, one could imagine a scenario whereby a single source at a certain frequency is sufficiently GW-loud to induce anisotropy and polarization signatures in the GWB, yet not be loud enough to be individually phase-resolved by template searches. This kind of ``emerging single source'' scenario has yet to be fully explored or characterized, but is critical to our understanding of how the first source resolved by PTAs will appear.

\subsection{Resolvable single sources}

This brings us to what will be the next major leap forward in PTA GW science--- the discovery of an individually resolved signal. Such a finding would not only be a huge breakthrough in its own right, but also affirm the astrophysical origin of the GWB. PTA collaborations have been searching for signals from individual SMBHBs for almost as long as the GWB. In fact one of the earliest successes of the PTA concept was the multi-messenger refutation (under certain assumptions) of the binary hypothesis for the core of radio-galaxy 3C66B, which exhibits variability on a $\sim$~$1.05$~year period \citep{2004A&A...426..379D}. The binary parameters implied by the variability model were such that a GW signal would have been easily detected, if not already apparent as a visible waveform in pulsar timing residuals \citep{2004ApJ...606..799J}. This system remains an important benchmark against which new PTA datasets and signal modeling approaches are tested \citep{2020ApJ...900..102A,2023ApJ...951L..28A,2024ApJ...963..144A,2025arXiv250820007C}.

There has been steady progress in modeling continuous GW (CGW) sources in PTAs, beginning from simple circular-binary Earth-term analyses \citep{2010PhRvD..81j4008S}, building toward eccentric binary models \citep{2016ApJ...817...70T,2020PhRvD.101d3022S,2023CQGra..40o5014S,2024ApJ...963..144A}, and incorporating the pulsar term ``echo'' of the emitting system's past state \citep{2010arXiv1008.1782C,2011MNRAS.414.3251L,2013CQGra..30v4004E,2022PhRvD.105l2003B}. The latter effect is analytically straightforward to model, but poses a computationally challenging MCMC sampling problem in Bayesian analyses. The addition of pulsar-term effects to a CGW search should only require an additional variable per pulsar denoting its distance, $L$, from Earth (or the Solar System barycenter). As discussed earlier in \S\ref{sec3}, the lag time between pulsar and Earth terms is $L(1-\cos\mu)$, where $\mu$ is the angular separation between GW origin and a pulsar. This can be of order $10^3$ years for $\sim$~kpc distant pulsars, implying potentially significant binary evolution for high chirp mass or high frequency sources. The trouble comes from evolving the binary orbital phase back by that amount of time; the PTA likelihood is highly oscillatory in sky location and pulsar distance, with numerous secondary maxima, unless this phase is simply treated as a nuisance parameter to be marginalized over per pulsar \citep[see, e.g.,][and subsequent works]{2014PhRvD..90j4028T}.

PTA analysis techniques for CGW searches have become mature enough through various optimizations to analyze modern datasets very quickly \citep[i.e.,][]{2022PhRvD.105l2003B}. Yet there are some trade-offs to address. Rapid Bayesian searches currently assume a CURN as a proxy for the Hellings-Downs-correlated GWB, after which posterior reweighting is needed \citep{2023PhRvD.107h4045H}; this will become ever more inefficient as a CGW gets more significant in data. Moreover, a PTA ``global fit'' (akin to what is envisioned for the LISA mission \cite[e.g.,][]{2025PhRvD.111b4060K,2023PhRvD.107f3004L,2025PhRvD.111j3014D,2024PhRvD.110b4005S}) has yet to be developed, but is crucially needed for transdimensional multi-CGW and GWB characterization campaigns. 

A key goal of CGW searches is localizing signals well enough to isolate the host galaxy, or even find electromagnetic counterparts. The latter could be possible if gas is present, where various observable signatures of binary-disk dynamics may render brightness modulations on timescales related to the binary orbit. Much work has been directed toward hydrodynamical simulations of this scenario, modeling the electromagnetic signatures arising from a disk-embedded SMBHB \citep[e.g,][and references therein]{2023arXiv231016896D}, assessing the prospects for overlapping demographics of SMBHBs detectable in both EM and GWs \citep{2019MNRAS.485.1579K,2022MNRAS.510.5929C,2025arXiv250821510C,2024A&A...691A.250C,2025arXiv250401074T} and even the potential for joint-likelihood multi-messenger searches \citep{2021ApJ...921..178L,2025arXiv250820007C,CharisiEtAl_JointLikelihood_inprep_2025}. Localizing the SMBHB's host galaxy will be a challenge, as the directional sensitivity of a PTA is limited by the broad antenna responses of each Earth-pulsar system. If PTAs were able to fully harness the power of pulsar-term information by knowing pulsar distances to within approximately a GW wavelength, PTA sky localization would be extraordinarily precise \citep{2012PhRvD..86l4028B,2011MNRAS.414.3251L,2010arXiv1008.1782C,2025arXiv250602819K,2023PhRvD.108l3535K}. Yet this distance precision eludes current efforts, leaving the expected localization areas for threshold CGW detections at $\sim10^2-10^3$~deg$^2$ \citep{2010PhRvD..81j4008S,2016ApJ...817...70T,2019MNRAS.485..248G,2024ApJ...976..129P,2025arXiv250401074T}, in which $\sim10^2-10^4$ plausible galaxies may lie. However, there are already efforts underway to develop filtering and probabilistic ranking criteria of candidate host galaxies in these credible sky regions \citep{2019MNRAS.485..248G,2024ApJ...976..129P,2024ApJ...961...34B,2024ApJ...977..265B,2025arXiv250421145H,2025arXiv250401074T}.

Through forecasting and simulation studies, we anticipate that the first SMBHB individually resolved by PTAs will typically be of the order $\mathcal{M}\sim 10^9-10^{9.5}\,M_\odot$, $z\lesssim 1$, with equal-mass systems preferred \citep{2009MNRAS.394.2255S,2015MNRAS.451.2417R,2018MNRAS.477..964K,2025CQGra..42b5021C,2025ApJ...988..222G}. These properties balance the underlying occurrence distribution of SMBHBs in the Universe---favoring low-mass BHs and a redshift distribution peaking further away---with PTA detector selection effects that favor nearby, equal-mass massive binary systems for their GW loudness. There are reasonable prospects for a small number of these systems to be embedded in disks that may give rise to electromagnetic counterpart signatures to the emitted GW signals. Time will tell, as SMBHB candidate systems are being found regularly through quasar lightcurve periodicity searches in time-domain survey datasets. The Large Scale Survey of Space \& Time (LSST) of the Vera C. Rubin Observatory is expected to find $\sim10^3-10^5$ more over its ten-year campaign, of which several hundreds may be true binary systems \citep[e.g.,][]{2025arXiv250821510C,2019MNRAS.485.1579K,2022ApJ...936...89W,2024ApJ...965...34D,2021MNRAS.508.2524K}. PTA collaborations are already carrying out targeted GW search campaigns on such candidates, albeit with currently inconclusive results. Yet these set a benchmark for the future--- if GW emission is confidently disfavored from such a target, the binary hypothesis is thus disproven, leaving the explanation as some other intrinsic brightness variability mechanism.

Forecasting the SMBHB detection timeline is challenging. More so than the GWB was, which is a persistent stochastic process, likely arising from the entire SMBHB population. But within the extreme, high-mass tail of the black-hole mass distribution, and in the relatively local Universe probed by the PTA sensitivity horizon, we are dealing with rather small-number Poisson statistics. Predictions range from detection within the next several years \citep[e.g.,][]{2017NatAs...1..886M}, to decades in the future \citep[e.g.,][]{2024ApJ...965..164G,2025ApJ...988..222G}. Prospects within the next ten years are reasonable, but it seems unlikely that PTAs will see a deluge of regular detections. Nevertheless, a handful of SMBHBs resolved out of, and resounding above, a stochastic GW background composed of a vast number of other binaries, make PTAs a unique experiment capable of probing the dynamics and demographics of the most massive compact objects in the Universe.

\subsection{New facilities and new horizons}

With the $2020$ loss of the Arecibo radio telescope, NANOGrav has relied on the Green Bank Telescope for much of its observations, complemented by CHIME (the Canadian Hydrogen Intensity Mapping Experiment; \cite{2021ApJS..255....5C}) and with a small number of observations at the Very Large Array. The emergence of newer PTA collaborations is incredibly exciting, including the Indian PTA that primarily uses the upgraded Giant Metrewave Radio Telescope \citep{2022PASA...39...53T}, the MeerKAT PTA using the MeerKAT array \citep{2023MNRAS.519.3976M} in South Africa (a precursor mission to the Square Kilometer Array, SKA; \cite{2015aska.confE..37J}), and the Chinese PTA using the Five Hundred Meter Aperture Spherical Telescope \citep{2025A&A...699A.165C}. In North America, prospects look incredibly good for a new facility currently called DSA-2000\footnote{\href{https://www.deepsynoptic.org/}{https://www.deepsynoptic.org/}} \citep{2019BAAS...51g.255H}; this will be $2000\times 5~\mathrm{m}$ radio dishes (or potentially fewer and larger) operating as an array in Nevada. If all goes well, this may become NANOGrav's new primary observational instrument in the late 2020s, making up $25\%$ of DSA-2000's observing portfolio. This could allow NANOGrav's array to expand to $\sim200$ MSPs, which, when complemented by the SKA and other facilities around the world, makes the future of PTA GW astronomy appear very bright.

PTAs are at once a mature discipline, and yet a field in the first flush of youth. The observations rely on the almost $60$-year old discipline of pulsar astronomy, but the first evidence of the low-frequency stochastic gravitational-wave background is still fresh. Techniques continue to be refined, assumptions revised, datasets lengthened, and data-analysis techniques necessarily made faster and more efficient. The near future should bring a fresh reaffirmation of this evidence at even higher significance through the International Pulsar Timing Array Data Release $3$, a truly global collaborative hunt for nanohertz frequency gravitational waves. 









\subsection*{Funding}
The author is a member of the NANOGrav collaboration, which receives support from NSF Physics Frontiers Center award number 1430284 and 2020265. The author is also grateful for support from grants NSF AST-2307719, and an NSF CAREER \#2146016. The author is supported by a Vanderbilt University Chancellor's Faculty Fellowship. 

\subsection*{Author contributions}
S.R.T was the sole contributer to the preparation and writing of this manuscript.

\subsection*{Data availability} 
No datasets were generated or analyzed during the preparation of this manuscript.

\section*{Declarations}

\subsection*{Competing interests}
The author has no relevant financial or non-financial interests to disclose.

\subsection*{Ethics declaration}
Not applicable.

\bibliography{sn-bibliography}

\end{document}